\documentclass[%
 reprint,
 amsmath,amssymb,
 aps,
]{revtex4-2}

\usepackage{graphicx}% Include figure files
\usepackage{dcolumn}% Align table columns on decimal point
\usepackage{bm}% bold math
\usepackage{graphicx}%
\usepackage{multirow}%
\usepackage{amsmath,amssymb,amsfonts}%
\usepackage{amsthm}%
\usepackage{mathrsfs}%
\usepackage[title]{appendix}%
\usepackage[dvipsnames]{xcolor}
\begin{document}

\preprint{APS/123-QED}

\title{Change point detection in ERA5 ground temperature time series}% Force line breaks with \\
%\thanks{A footnote to the article title}%

\author{Fatemeh Aghaei A.}
\email{f.aghaei.a@gmail.com}
 \affiliation{Max-Planck Institute for the Physics of Complex Systems, Nöthnitzer 38, 01187, Dresden, Germany}%Lines break automatically or can be forced with \\

\author{Ewan T. Phillips}%
 %\email{Second.Author@institution.edu}
\affiliation{Max-Planck Institute for the Physics of Complex Systems, Nöthnitzer 38, 01187, Dresden, Germany}%

\author{Holger Kantz}%
 %\email{Second.Author@institution.edu}
\affiliation{Max-Planck Institute for the Physics of Complex Systems, Nöthnitzer 38, 01187, Dresden, Germany}%

\date{\today}% It is always \today, today,
             %  but any date may be explicitly specified

\begin{abstract}
We analyze the ERA5 reanalysis 2-meter temperature time series
   on all land 
  grid points using change point analysis. We fit two linear
  slopes to the data with the constraint that they merge at the point in time 
  where the slope changes. We compare such fits to a standard
    linear regression in two ways: We use Akaike's and the
      Bayesian 
      information 
    criteria for model selection, and we test against the null hypothesis of
    no change of the trend value. For those grid points where the
  dual linear fit is superior, we construct maps of the time when the trend
  changes, and of the warming trends in both time intervals. In doing so, we
  indentify areas where warming speeds up, but find as well areas where
  warming slows down.
  We thereby contribute to the characterization of local effects of climate
  change. We find that many grid points exhibit a
  change to a much stronger warming trend around the years 1980$\pm$10. This raises
  the question of whether the climate system has already passed some 
  tipping point.\\

\end{abstract}

\maketitle

\section{Introduction}\label{sec1}
Climate change has been a significant concern of the scientific community
since at least the mid-1980s, as highlighted by the formation of the IPCC in
1988 and its first scientific report in 1990\cite{IPCC}. In fact,
Arrhenius had as early
as 1896 predicted the effect of carbon dioxide on the radiation budget of
Earth\cite{Arrhenius}. Reconstrutions of the global mean surface temperature
of the Earth performed by different research
organizations\cite{globalmeantemp} all agree in showing a clear 
warming trend starting at the latest in 1975, while an increase beyond the
pre-industrial level may be present since the early 20th century.

By contrast, local temperature measurements show much more complex patterns
which deserve detailed analysis, since regional or even local climate
change patterns are of utmost relevance for a maximally efficient mitigation
strategy. Changes in temperatures affect human health, the selection of crops
for sustainable agriculture, forestry, and tourism, but also the local water cycle
and even transportation. Only with a good understanding of the expected
changes of local climate, these issues can be addressed.

There are many concerns about tipping points in the climate system. The
concept of tipping describes a feedback loop which, once it has been
triggered, cannot be ``switched off'' by small interventions any
more. Examples include the melting of permafrost ground and massive release of
methane into the atmosphere, the disappearance of the Arctic sea ice with a
lowering of the ice albedo effect in the polar region,
and massive CO$_2$ release due to wildfires as
a consequence of changes in the water cycle in the wake of
warming \cite{Lentonetal}.  While
there is literature on \textit{predicting} upcoming tipping
events\cite{tipping_precursor, review1},  the issue of
\textit{detection of having passed a tipping point} is much less
explored in the literature. By the detection of change points in local warming
trends presented in this paper, we intend in particular to highlight those
years in which climate change has gained momentum. We consider these years to
be candidates for past tipping events.

Inhomogeneity of the warming trend as a function of latitude as
well as geographic location has attracted the attention of
many researchers \cite{Inhomogeneity1, Inhomogeneity2, Inhomogeneity3}. Of
particular interest is the warming trend of Antarctica, which has 
found to be non-existent or even negative in some studies \cite{Antarctica1,
  Antarctica2, Antarctica3}, as 
well as other unexpected regional cooling effects \cite{specialregion1}. Regional warming patterns have been shown to possess relevant impacts of
various kinds on, e.g.,
the water cycle, vegetation, health, and other elements of
the biosphere \cite{impact1, impact2}.

In this paper, we focus on re-analysis temperature data
from the ERA5 project from 1950 to
March 2021. For being able to analyze local time series covering the
  whole land mass of the globe in 
Sec.\ref{sec:local}, we use as gridded data the 2-meter above-ground daily mean
air temperature time series of the ERA5 re-analysis project
with $1^{\circ}\times 1^{\circ}$ resolution\cite{ERA5, ERA5_2}. \\
ERA5 is a comprehensive climate reanalysis dataset produced by the European Centre for Medium-Range Weather Forecasts (ECMWF) under the Copernicus Climate Change Service (C3S). It provides hourly estimates of atmospheric, land, and oceanic climate variables on a 0.25° × 0.25° grid (approximately 30km spatial resolution) from 1940 to the present on 37 pressure levels. In addition, there is a higher-resolution product for the land surface, called ERA5-Land, which includes climate variables such as temperature, pressure, wind, humidity, and precipitation on a 0.1° × 0.1° grid (approximately 10km spatial resolution).
\\
ERA5 data are produced through a sophisticated process that combines historical observations with numerical weather prediction (NWP) models using data assimilation techniques. Data assimilation blends information from diverse observations (e.g., satellite data, weather stations, radiosondes) with model forecasts by adjusting the model state to minimize the differences between the observations and the model predictions\cite{era5_reanalysisProcess}.
\\
The core model used in ERA5 is the ECMWF’s Integrated Forecasting System (IFS). It simulates atmospheric processes using the fundamental laws of physics, including momentum equations for wind vectors, continuity equations for air density, and thermodynamic equations for temperature and pressure\cite{era5_reanalysisProcess}.
\\
By collecting observations from a variety of sources and combining them with the NWP model, the data assimilation technique can fill in missing data and appropriately weight the uncertainty of the estimates in regions or periods with sparse observations.
\\
Because ERA5 continuously reprocesses past data at hourly intervals, it provides a consistent long-term record from 1950 to the present.
\\
ERA5 datasets are available through the C3S Climate Data Store (CDS)\cite{ERA5} and from the ECMWF\cite{ECMWF}. In this work, we used annual data constructed by averaging over the higher-resolution data. Although one could aggregate the original hourly data to obtain annual values, ERA5 also provides post-processed daily statistics, which are available for download from the CDS via its web interface or API service\cite{ERA5_2}.
\\

For the analysis of the warming trend, we use annual averages of the daily
temperatures, so that we work with time series of 72 values at every grid point.
%we remove the individual seasonal cycle, so the resulting anomalies exhibit large fluctuations, with a standard deviation of about 4-6K
%(depending on geographical location), superimposed by natural climate
%variability in these 70 years of data.
Rather strong fluctuations related to natural climate variability almost 
mask the warming trend due to climate change, since the
global mean time-averaged temperature in 2023 
 was about 1.36K\cite{nasa} higher 
than in pre-industrial times. Taking into account the long-range
temporal correlations present in such time series \cite{Koscielny,KoscielnyBunde,Bundeetal}, which
introduce redundancy in the data, observed local trends are usually
statistically significant at the 95\% level only if they
exceed 0.20 K/decade on a
70-year long time series (details in Sec.\ref{sec:global}).

It is evident that local and global temperature changes are
not well described by a single linear trend over the full time span.
When considering the future we expect to see sigmoidal
  temperature curves which saturate on a new level. These could be characterized
  by the
  step height, by the time of the steepest increase, and the value of this
  slope. However, currently we are still in a situation where temperatures
  increase. For local observations, no simple functional form for the time dependence
  of temperture has been proposed. Smoothing observation data by filters
  allows clearer observations of the warming, by remains qualitative since
  there are no easily interpretable parameters, hence filtering data yields a
  non-parametric model.   We instead perform fits
  of two linear slopes which merge at a change point.
  In doing so, we can extract relevant
  information about regional to local climate change in terms of \textit{when} the
  warming trend has changed and the values of \textit{how} the trend
  has changed, so our model is parametric.
  We compare this model to the null hypothesis of a single trend
  value (i.e., standard linear regression) in terms of model
  selection criteria and in terms of rejection of the null hypotheses.

\begin{figure}
    \centerline{\includegraphics[scale=0.75]{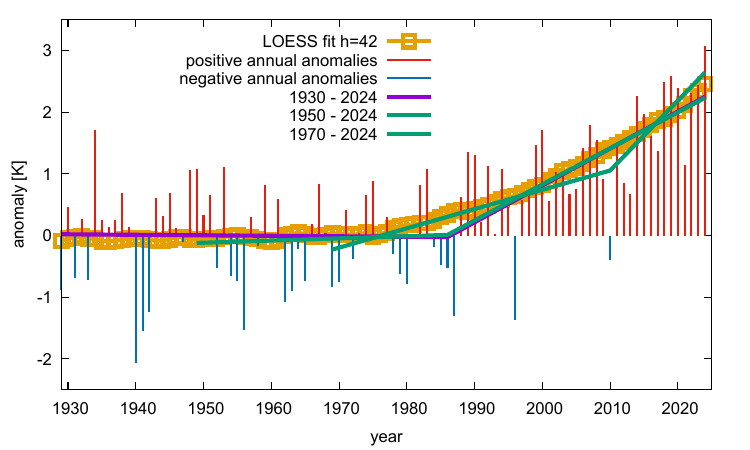}}
    \caption{\label{fig:Potsdam}
 The anomalies of annual mean temperatures of
      Potsdam, Germany (raw data via Deutscher Wetterdienst \cite{CDC1})
      with respect to the period
      1961-1990, with a LOESS fit (yellow symbols) and our continuous
      two-slopes fits, performed on different subsets of the data. The fit done
      on the years 1970-2024 is not significant because the
      Bayesian Information Criterion favours a single linear regression.}
    \end{figure}

 As a motivation, we show in Fig.\ref{fig:Potsdam} the annual temperature
  annomalies measured in Potsdam, Germany, where as reference we subtracted
  the average temperature during the period from 1961-1990. Following the
  report of the German Weather Service DWD \cite{report_2024_DWD}, we perform
  a smoothing of these data by a LOESS filter with a bandwidth of 42 years,
  the result of which indicates a strong increase of the warming during
  the
  period of 1975-1995 (as a note, this depends partly  on the bandwith of the
  LOESS kernel). We include the results of our continuous two-slopes model
  fitting it to three different time intervals of the \textit{raw data}.
  While the periods from
  1930 till today and from 1950 till today agree very well and identify a
  change point in the year 1986, doing the same for 1970 till today results in
  a different change point and different slopes. However, using the Bayesian
  Information criterion BIC for model selection (details in Sec.2),
  we find that the change point
  in this latter interval is insignificant and a single linear regression is
  the better model. This exemplifies a certain robustness of our approach
  which yields a definite year of change and the temperature trends before
  and after the change together with a significance test for the existence of
  such a change point. In the main part of the paper we will explain our method
  in detail and perform tests on numerically generated data with and without
  change points.

The results of this paper are maps of the globe showing when a change of the
warming trend 
occurred, if such a change is statistically significant, and to show how the
warming trend has changed, for every grid point on land on a $1^o\times 1^o$ gridded temperature
data set.

In section \ref{sec:methods+data}, we present the methods of analysis and the data sets. 
In Section \ref{sec:statistics}, we discuss the accuracy of our method using
synthetically generated data and we introduce the statistical tests for model
selection and for testing against a null-hypotheis. In Section \ref{sec:global}, we analyze the global mean surface temperature of the Earth in
terms of a time-local warming trend, together with its error bars, highlighting
both the overall warming but also the natural climate
variability. We extend the global observed data series into
  the future by using climate projection data for 3 different scenarios. This
  provides evidence that a continuous two-slopes model is a reasonable and yet
  parsimonious model with easily interpretable parameters.
In Section \ref{sec:local}, we analyze the grid point time series in view of change points of the 
warming trend and discuss their variations across the globe.
In Section 6, we interpret these findings and discuss the limitations of this
method.

\section{Methods and Data\label{sec:methods+data}}
While climate projections using climate models
provide a good outlook on warming under different
emission scenarios, they do not predict a specific functional dependence
$T(t)$ for the global temperature $T$ on time $t$ which one could fit to
observation data. If humankind succeeds in
controlling climate change, one would expect a sigmoid function for $T(t)$,
with a saturation value hopefully below 2K above the pre-industrial
times. Unfortunately, our insight into the data does not (yet)
show any signal of
\textit{slowing down} of global warming so that a fit of a sigmoid function to
observed data 
does not make sense. Instead, in the majority of grid points, we find an
\textit{acceleration} of warming in the past one to two decades.

The most
parsimonious fit to the data in this situation is a dual-linear fit of two
slopes with a change point, where the two slopes should merge. This can be
interpreted as an approximation to the first part of a sigmoid where its
steepness still increases. We enforce continuity of the fit
at the time of change because of the continuity of all natural processes in
particular in 
the radiation budget of the globe. More importantly,
accepting an additional
jump at the change point can lead to misleading results. In numerical
tests of fits with two slopes without continuity to data with a single trend,
with about 20\% probability such a fit produces at least one
negative slope with a large
positive jump in between, and in rare cases even 2 negative slopes with an
even bigger 
jump which ensures that the mean value on the second segment is larger
than that of the first, as a consequence of the positive trend.

Our optimization problem means to first finding the
optimal slopes on both segments of the data under the constraint that at a
pre-determined time, the two linear segments merge, and in a second step we
optimize this time of change by minimizing the overall root-mean-squared (RMS)
error
of this fit with respect to the used change point. The first part is a linear
optimization problem that can be treated analytically, see Appendix A.
We hence have closed formulae for the two slopes and the two intercepts
of the dual-linear fit as a function of the observed data and the chosen time
of change, which can be easily evaluated numerically. We then let the time of
change run through all years starting in 1960 and ending in 2010, because for
robustness we require that each segment has at least 10 data points to be
reliably fitted with a linear function. We select
as the best fit the one where the total RMS-error is minimal. We also look
into relative minima which are still close to the absolute one, in order to
better understand the timing of the change. For verification of the
statistical significance of such a fit, we will compare it to a standard linear
regression with a single slope for the full time span (null hypothesis)
by help of Akaike's and the Bayesian information criterion, and we will
perform a 
statistical test against the acceptance of the null hypothesis with a 95\% confidence
level.

There is a large body of
literature on change point detection,  largely discussed in
the recent review \cite{review_change_points}.
The method employed here is specifically apt
for the analysis of a changing warming trend. In other words,
we model a change of the time derivative of temperature, but not a jump in
temperature itself.

While this analysis may be done
on the daily anomalies ($T(t)$ subtracted by the local seasonal cycle), one
can speed up the analysis tremendously by simply considering the annual mean
values of the raw data. Taking the annual mean  
averages out the seasonal cycle so
that no anomalies are needed, and it reduces the number of data items in the
constraint linear fit by a fector of 365 with a gain in numerical stability.
We veryfied carefully that the results of the analysis on daily anomalies and on
annual mean values agree with high precision.

% **********************************************
\section{Statistical uncertainty of the dual-linear fit\label{sec:statistics}}

Before applying the dual-linear fit to the temperature time series of the grid
points, we will discuss here how we access the statistical significance of our
results. There are 2 distinct issues: (a) if the data follow our model of two
linear segments merging continuously with superimposed fluctuations, how
accuratly will our fit identify the time of change? and (b) given some
arbitrary data, how do we verify that our dual-linear fit is an appropriate
model for the data? For (b) we compare our model to a simple linear regression
and use two approaches to decide which model is better, namely by studying the
chance of over-fitting, and by rejecting a null-hypothesis.

We start with discussing issue (a). To this end, we
generate an ensemble with 1000 members
of artificial temperature time series for a 70-year time interval
with a change point at year 35 when we switch from a stationary process to
one with a trend of 4K/century.
To simulate the stochasticity we add annual mean value
anomalies from a Gaussian distribution with a standard deviation of 0.45K, a
typical value extracted from the Potsdam temperature time series
\cite{Potsdam_data}.

\begin{figure*}[ht]
\centering
\includegraphics[scale=0.5]{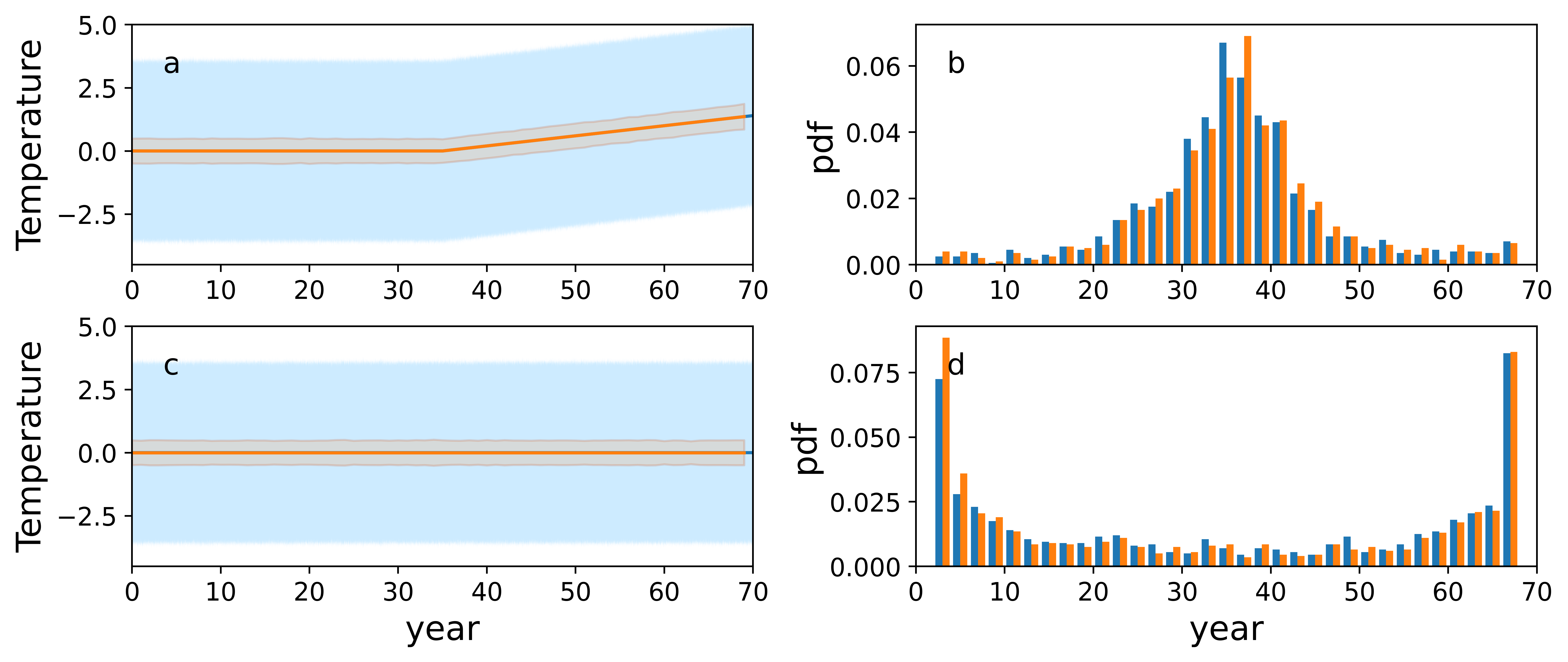}
\caption{Deterministic trend (orange) and standard deviation (light blue (daily) and
  light orange (annual) shadows)
  of artificial time series
  covering 70 years for (a) two different 
  deterministic trends merging in year 35 and (c) no change point (no trend),
  superimposed by white noise. \\
  The histograms of the detected change
  points obtained by the dual-fit method for ensembles of 1000 time series, (b) where the true
  the change point is in the year 35, (d) without change point in the data
  model.
Blue and orange in the histograms and the standard deviations refer to daily data (blue) and annual averages
(orange).}
\label{Fig_synthetic}
\end{figure*}

Figure \ref{Fig_synthetic} (a-b) shows the results of the dual-linear for
synthetic data set. We repeated this analysis for synthetic data with
fluctuations generated by a long-range correlated ARFIMA(p=0, d, q=0) process
(autoregressive fractionally integrated moving average, where p is the order
of the autoregressive model, d is the degree of differencing, and q is the
order of the moving-average model). With H = d + 1/2 = 0.65 and 0.8, we found
a distribution of detected change points very similar to that of white noise
anomalies. 

While the ground truth in these data is that the change point appears in the
year 35, due to the superimposed randomness the optimal fits can identify
different years as change points. This depends on the signal-to-noise ratio
given by the standard deviation of the fluctuations and the total systematic
change due to the trend on the full time interval. For realistic values of both,
the change point is detected correctly with $50\%$ probability inside the
interval 
$\pm 5$ years, and with $80\%$ probability inside $\pm 8$ years. For smaller standard deviation of the random
  fluctuations or for larger trend values the distribution of detected times
  of change concentrates much more around the truth, but the performance shown
  here it typical of real temperature data.
However, for some realizations of the stochastic perturbations, the
fit identifies erroneously change points at times quite far away from
the truth. 
The possibility of such outliers has to be taken into account when
we use this analysis on about 20000 time series on all land grid points
of the Earth.
The significant role of stochastic perturbations in detecting tipping points
raises the question of what results from dual-linear fit analysis when there is
no change point in deterministic dynamics. To explore this, we repeat our analysis
for data without change point, where stationary fluctuations are
generated by the same Gaussian distributions. 
In this case we obtain a pathological result that the change points are
concentrated mainly on the first and last years, see
 Fig.\ref{Fig_synthetic}(c-d). This finding challenges the notion of a genuine
 change point. Indeed, the analysis shown in Fig.\ref{Fig_synthetic} always
 assumes a change point to be hidden in the data. In order to suppress
 meaningless results, we therefore will modify our analysis in two ways:
 First, we only test for change points that are at least 10 years away from
 the beginning and from the end of the time series, and we will in addition
 compare the dual-linear fit to a single slope fit by help of information
 criteria and a hypothesis test, i.e., will will intoduce a model-selection
 step (b). 

 The dual linear model has 4 fitting parameters and therefore the ability to
 fit data with lower root mean squared (RMS) 
 error values than the single linear model. Hence, RMSE is not an appropriate
 criterion for model selection. Instead, scores like the Akaike Information
 Criterion (AIC) and Bayesian Information Criterion (BIC) offer a more
 comprehensive evaluation. These metrics assess the balance between goodness
 of fit (measured by RMS error) and model simplicity (measured by the number
 of parameters). A lower criterion value indicates a better trade-off between
 simplicity and fit goodness \cite{AIC, BIC, ModelSelection}.
 
 Originally proposed based on maximum likelihood, AIC can be expressed as a
 function of the residual sum of squares (RSS), $RMSE = \sqrt{RSS/N}$, for
 Gaussian distributed residuals. BIC is akin to AIC however it takes into
 account the 
 number of data points not only in the goodness of fit but also in the
 complexity 
 term. These criteria provide an insight of whether the accuracy a model
 achieves justifies its complexity \cite{ModelSelection2}.
 For a time series of length $N$ and for a model with $k$ fit parameters,
 AIC and BIC read:

\begin{eqnarray}
\mbox{AIC} &=& c(N) + N\log(RSS/N) +2k\label{Eq_AIC}\\
\mbox{BIC} &=& N\log(RSS/N) + k\log(N)\label{Eq_BIC}
\end{eqnarray}

We assessed both the AIC and BIC metrics for our synthetic data, comparing the
performance of single-segment and double-segment models. Our observation
suggests that BIC is more reliable, given that the dual-linear model
statistically tends to be favored by AIC, regardless of whether it is applied
to single-segment or double-segment data. 
\\

\begin{figure}[ht]
\centering
\includegraphics[scale=0.65]{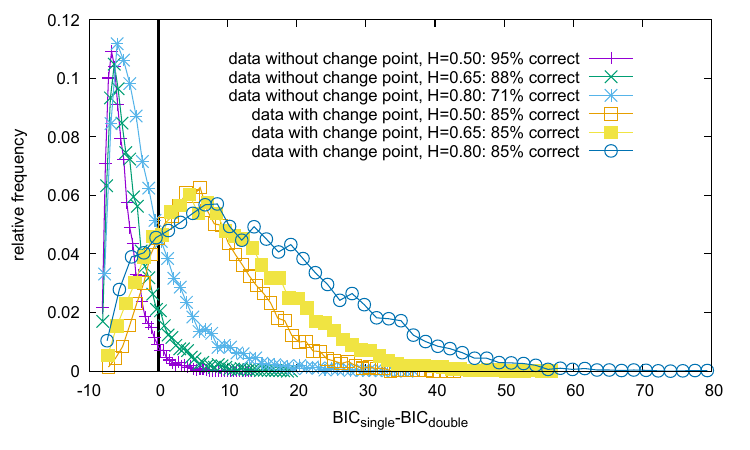}
\caption{ The distribution of
 BIC$_{single-line~model}$ - BIC$_{dual-line~model}$ for 10000 ensembles for two cases, without change point and with change point, for time series with Gaussian noise and long-range correlated noise. Model
 selection due to BIC means to take the model with the smaller BIC value.
\label{BIC_synthetic}}
\end{figure}

Fig. \ref{BIC_synthetic}, which represents the distribution of $\Delta BIC = BIC _{single-line~model} - BIC_{dual-line~model}$ for ensembles of time series with Gaussian fluctuations (H = 0.5) and long-range correlated fluctuations (H = 0.65, 0.80), demonstrates BIC's ability to distinguish dual-segment cases from single-segment ones, given that model selection based on BIC favors the model with the smaller BIC value. 
As expected in the cases without a change point, the distribution is concentrated in negative $\Delta BIC$, whereas for two-segment ensembles, it is centered in positive $\Delta BIC$. The figure indicates that in two-segment cases, the criterion reliably selects the correct model with 85\% accuracy, regardless of the time series' correlation strength. On the other hand, in single-segment cases, the model's performance depends on the correlation strength, weakening as the correlation strength increases. 
Since temperature time series anomalies exhibit a correlation strength around 
H=0.65 \cite{Johannes}, which corresponds to an accuracy of 88\%, we can infer that the BIC-based model selection remains highly effective in distinguishing structural changes in realistic climate data.

It has been proven that in the limit of
$N\to\infty$, BIC will correctly identify the data generating model
\cite{DingTarokhYang}, while AIC is known to be asymptotically equivalent to
leave one out cross validation\cite{Stone}.

A very different philosophy of model selection is hypothesis testing. In our
setting, this means that when performing the dual-linear fit we can (or cannot) reject the null hypothesis of the absence of a change point with a given
confidence. For this purpose, we consider the absolute difference of the two fitted
slopes as a test statistic, $s=|a_1-a_2|$. We then generate the distribution
of $s$ for a large random ensemble of data under the null-hypothesis, i.e.,
generated without a change of the trend, and identify the one sided confidence
interval of this distribution. If a dual-linear fit to data with unknown trend
produces a value of $s$ outside this interval, we can reject the
null-hypothesis with 95\% confidence. As a technical complication, the
$s$-values under the null-hypothesis tend to be the larger, the more the time
of change is found towards the beginning or end of the data set. Hence one has
to determine the 95\% confidence interval conditional to the time of change
found. In the application to the grid point temperature time series, we will 
perform this test of statistical significance. We find a very good agreement
 between statistical significance and superiority of the BIC value so that
 both
 criteria lead almost always to the same conclusions.

 To summarize this section, for synthetic data with a single change point
 and slopes and fluctuations which are chosen to represent the grid point
 temperature time series,
 in about 80\% of individual data sets the change point is detected
 in an interval of less than 8 years around the true value. This might appear
 to be not very precise, but is due to a low signal-to-noise ratio:
 the amplitude of the random fluctuations due to 
 natural climate variability (standard deviation $\approx$ 0.5K) are
 of the same 
 order of magnitude  as the total warming with respect to the pre-industrial
 times ($\approx$ 1.5K).

\section{Analysis of the global mean temperature\label{sec:global}}
Before we discuss the results obtained for individual grid point time series, we
analyze the series of the global average land temperature of Earth from
1950 to 2021. The
global temperature is obtained as the weighted average of the temperatures of
all land grid points taking into account their corresponding areas which is a
function of the latitude\cite{weighting_grid_points}.

\begin{figure*}
\centering
\includegraphics[width=\linewidth]{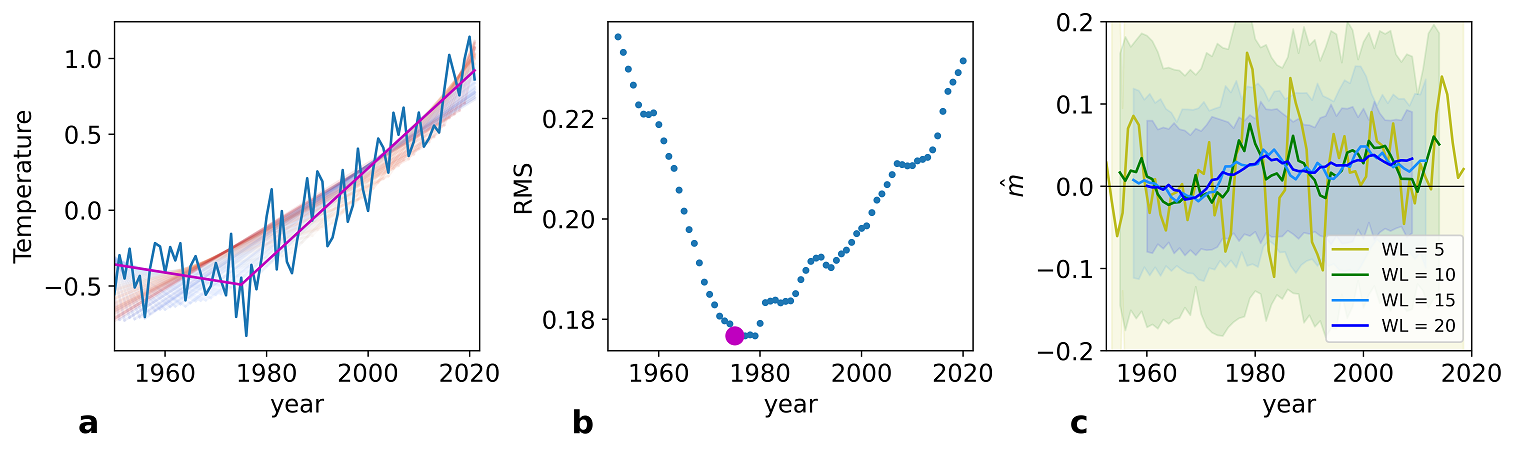}
\caption{
Applying the dual-linear fit to the global temperature time series. (a) The
blue curve is the average temperature of the land grid points of ERA5 data.
The shadow behind the time series illustrates all possible double
linear fits where the optimal one is specified by magenta color. (b) Root Mean
square error of the dual-linear fit as a function of the year used  
as a change point. (c) 
Time-local slopes (warming trends) calculated on moving windows of length 5, 10, 15, and 20
years are shown by solid lines as function of the year in the center of the window.
The shadows with the corresponding color show the error bars 
$\sigma[\hat{m}]/2$ for the estimated trend values.
(The uncertainties of the trend estimates on 5 year windows exceed the range of
the $y$-axis).
\label{global_data}}
\end{figure*}

The annually averaged time series of the
global temperature is presented in Fig.\ref{global_data}. Our dual-linear fit
method identifies a change point between 1976 and 1980 when the slope changed from $-0.27
K/$century to $3.03 K/$century which is $1.80 K/$century on average. This
outcome states that the global mean temperature was decreasing slowly while in
the 1970s it changed to a strongly warming phase. The RMSE shown in
Fig.\ref{global_data}(b) as function of the time of trend change has an approximate 'V' shape so that there are no
further plausible candidates for times of change.
We repeated the same analysis
with the global temperature series including both land and
oceans supplied by NOAA\cite{NOAA_global}. This analysis also revealed a change point between 1976 and 1980, however, it indicates that the globe had
a weak warming trend of $0.36 K/$century before and a much faster warming of
$1.98 K/$century which is $1.4 K/$century on average. 
Evidently, the high heat capacity of the oceans leads to different trend
values compared to those obtained only from land data. 
In addition, our analysis of the monthly HadCRUT5, GISS, and UTA temperature
time series produces similar results (data from \cite{globalmeantemp}).

\begin{figure*}
\centering
\includegraphics[width=0.6\linewidth]{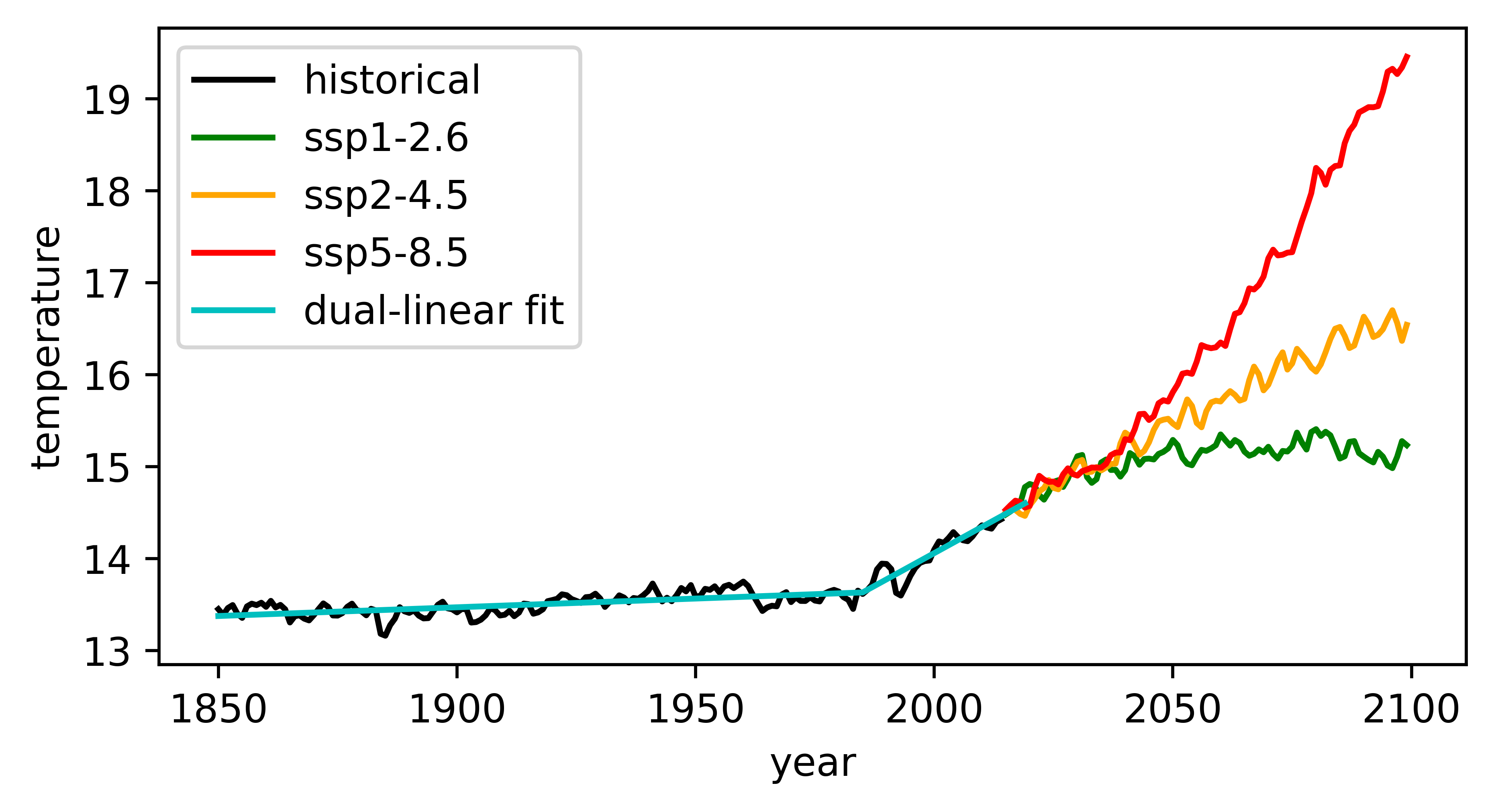}
\caption{
  The dual-linear fit applied to three different climate projections. The fit
  was done to data in the years 1850-2040, while data are shown farther into
  the future. Evidently, for ssp1-2.6 a dual linear fit on the whole time
  would not make sense since we see a sigmoidal behavior of the temperature curve.
\label{global_data_future}}
\end{figure*}

There is some discussion about
change points in the global mean surface temperature, see
e.g.\cite{Rahmstorf}, to which we do not want to contribute, since we are
interested in local time series. Let us stress, however, that any result on
change point analysis depends on the time span covered.
We therefore repeat
the analysis for climate projections extending into the future, for 3
different SSPs (Shared Socioeconomic Pathways) scenarios including SSP1-2.6 (Sustainability, Taking the Green Road), SSP2-4.5 (Middle of the Road) and SSP5-8.5 (Fossil-Fueled Development). The SSPs are future greenhouse gas concentration scenarios developed by the IPCC (Intergovernmental Panel on Climate Change). They incorporate varying assumptions about population growth, economic development, and climate policies, and are used to project how human activity may influence future greenhouse gas emissions, and consequently global temperature. The results in
Fig.\ref{global_data_future} shows that when restricting the fit to the time span from 1850 to 2045, all different projections lead to a time of change that matches that of global NOAA and ERA5 data. 

% Beyond the expected change point caused by global warming,
Natural climate
variability affects the global mean temperature on various time scales.
We illustrate this variability by fitting warming trends on overlapping moving windows for 5, 10, 15,
and 20 years, which are the typical time scales of variability in climate
systems. In addition, we calculate the uncertainty of the estimated trend
values taking into account the sample size, the short-range temporal
correlations and also the long-range temporal correlations which are present
in all temperature time series\cite{Koscielny,Iwanetal}.
Such long range correlations, represented by a Hurst exponent larger than 1/2, are able to
dramatically increase
the variance of the least squares trend estimator compared to a  white noise
signal. The standard deviation (square root of the variance) can be
used as the magnitude of the error bars for the estimated trend $\hat
m$. Given knowledge of the short and long range correlations, as well as the
assumption of Gaussianity, the variance can be calculated:
\begin{equation}
\sigma^2[\hat{m}]  \sim \sigma^2[T]~f(\phi, d)~N^{2d-3}.
\label{Eq_uncertainty_variance}
\end{equation}
Here, $\sigma^2[.]$ is the variance, $\hat{m}$ is the fitted slope, $T$ is the
temperature time series, $N$ is the number of data points in the time window,
$\phi\in(-1, 1)$  is the auto-regressive parameter representing short range
correlations, and $d\in(-0.5, 0.5)$ is
related to Hurst exponent $d=H-1/2$, and $f(\phi, d)$ is calculated
as follows: 

\begin{equation}\label{Eq_uncertainty_f}
\begin{split}
f (\phi, d) = &\frac{1+\phi}{(1-\phi)(2~{}_2{F}_1(1, d, 1-d, \phi)-1)} \\
&\times \frac{36(1-2d)~\Gamma(1-d)}{d(1+2d)(3+2d)~\Gamma(d)}.
\end{split}
\end{equation}
where ${}_2{F}_1(.)$ and $\Gamma(.)$ are the hypergeometric and Gamma
functions\cite{Iwanetal}.
For the data set of the global mean surface temperature obtained from ERA5
we determine  $d=0.29\pm 0.03$,
using detrended fluctuation analysis \cite{DFA}. We then apply the
Gr\"unwald-Letnikov derivative and obtain $\phi=0.87\pm 0.04$ from the lag-1
autocorrelations as in \cite{Johannes}.

Figure \ref{global_data} shows that the Earth experienced climate
variability on different time scales. Trends calculated on 5-year and 10-year
windows suffer from large error bars and fluctuate in magnitude and sign,
showing that climate change cannot be characterized on these time scales. On
intervals of 15-20 years, trends stabilize. A clear warming signal
evolved from the 1970s onward, and despite low amplitude fluctuations, the
trend values have remained positive since then. However, these values are
still, within the error bars, compatible with a stable climate. 

Figure \ref{global_data} also shows that there is no simple functional form
for the temperature change. A 
constant trend value during the full 72 years is as much of an
oversimplification as any other analytical curve. This motivates us to use our
two-slope model in order to further characterize climate change locally, where
all types of fluctuations have much larger amplitudes compared to the
trend than in Fig.\ref{global_data}, see Fig.\ref{fig:Potsdam}.
Clearly, one could generalize this method, using fit
functions with more than two linear
trends. We refrain from  from doing so, because the results of such fits are
more difficult to present, to compare, and to interpret, but we will discuss
the issue of model misspecification in the Conclusions.

\section{Change point analysis of grid-point time series and regional
  properties\label{sec:local}}
\begin{figure}
\centering
  \includegraphics[scale=0.6]{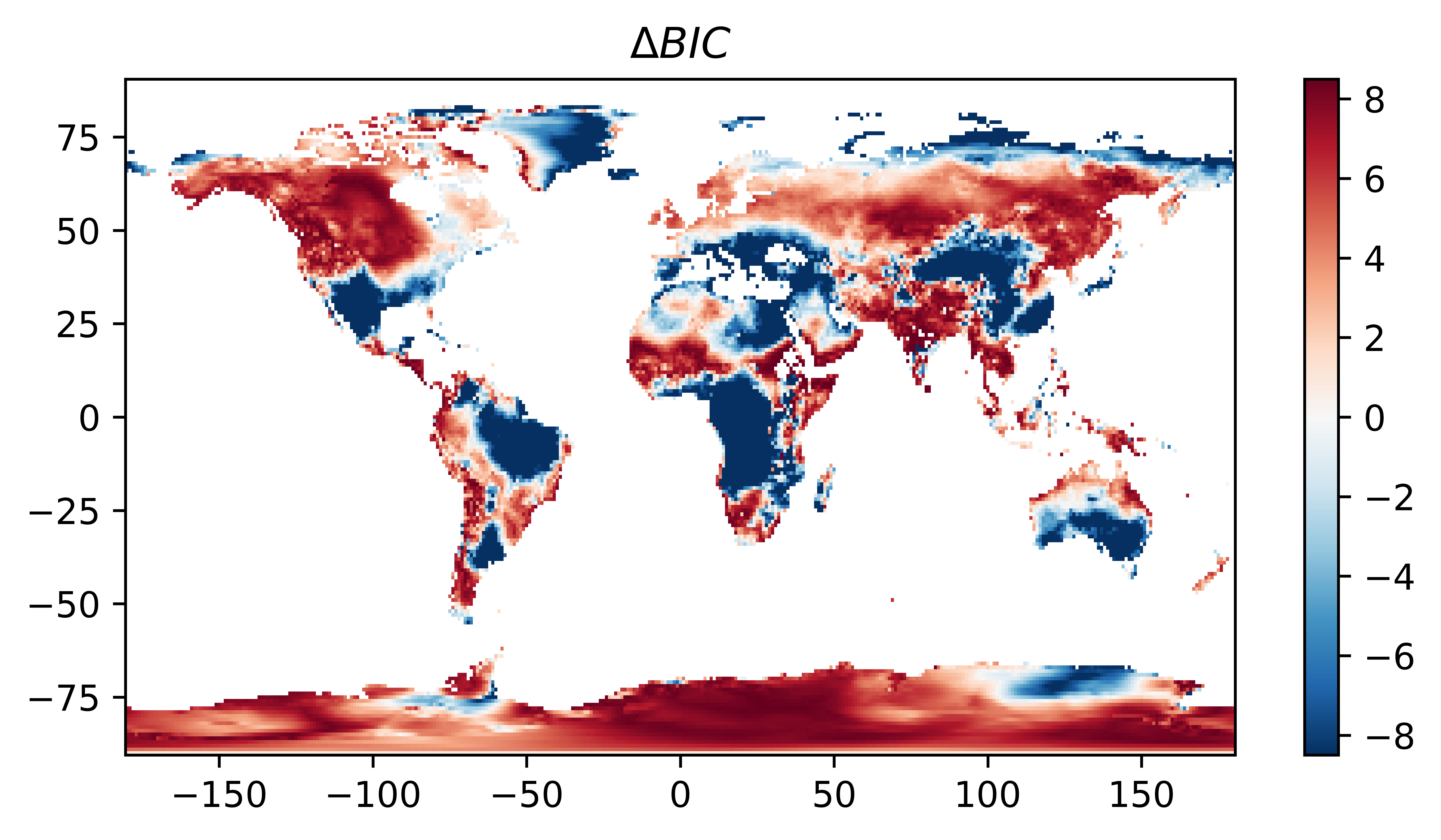}
  \caption{The difference between the Bayesian scores (BIC) for the dual linear
    model and single linear model,
    $\Delta BIC = BIC_{dual-linear~model} - BIC_{single-linear~model}$.}
     \label{dBIC}
\end{figure}

In this Section we analyze ERA5 2m above-ground temperature data at each individual grid
point on land of the $1^o\times 1^o$ degree data set\cite{ERA5_2}. The goal is to detect whether
such local data show a significant change of the warming trend, and if so,
when it occurred.  This means that we repeat the analysis 
shown in Fig.\ref{global_data} (a)-(b)
together with a single linear slope fit to the annually averaged local temperature
series from 1950 to 2021 and compare BICs of the
two models, as well as evaluate whether we can reject the null-hypothesis of
  the absence of a change point with more than 95\% confidence. Since the
statistical significance test and the BIC criterion agree on more than 94\% of
all grid points, we only show results using the BIC as model selection statistics.
In Fig. \ref{dBIC}
we show the differences between the BICs for the double linear
model and the single linear model, $\Delta \mbox{BIC} =
\mbox{BIC}_{dual-linear~model} - \mbox{BIC}_{single-linear~model}$
as a color coded map. A negative $\Delta \mbox{BIC}$, denoted by blue color,
signifies that the double linear model is superior to the
single linear one. Also, the light red spots
(slightly positive value of $\Delta \mbox{BIC}$) are debatable considering
what we 
observed from the numerical experiments that $\mbox{BIC}_{dual-linear~model}$
might 
get a slightly higher value than $\mbox{BIC}_{single-linear~model}$ even
though there 
are two slopes. On the other hand, the dark red colors indicate that 
the single linear fit is the preferable model. Our investigation shows that
when taking into account the area corresponding to each grid point, the
temperature in approximately $50\%$ of the global land area has experienced a
statistically significant change in trend over these years 
(in terms of 95\% confidence).
Regions where
the single-linear model is preferable can be found in different parts of the
globe including central Asia, North America, west of Africa, western border of
South America, but more than $44\%$ of them are located in Antarctica, which is
less than $10\%$ of the total area. Therefore, here we can conclude that our
dual linear fit and the concept of its change point is a meaningful analysis
for the temperature dynamics of the globe.

\begin{figure}
\centering
  \includegraphics[scale=0.6]{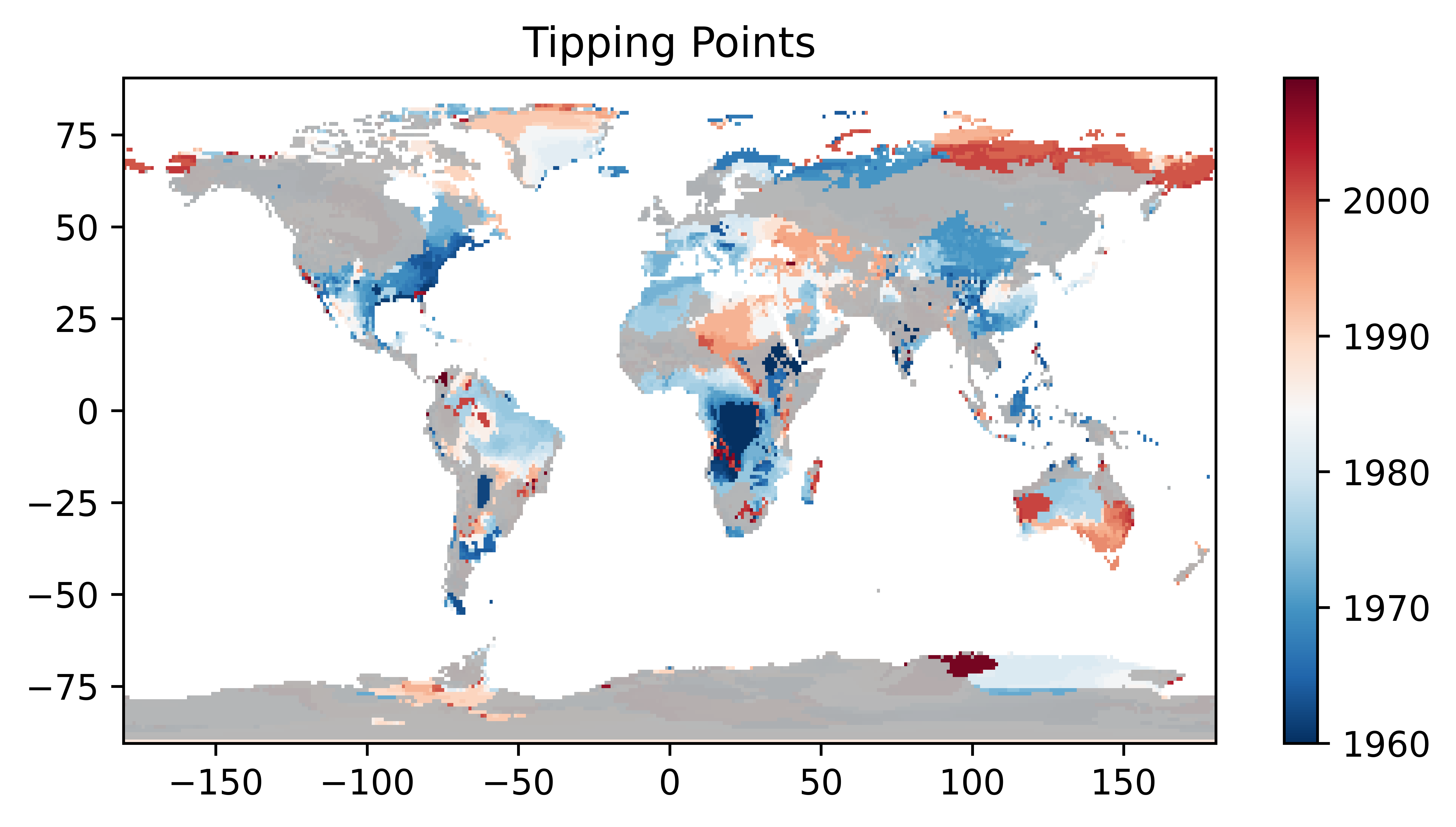}
  \caption{
    The years of change detected by the optimal dual-linear fit method in
    color code. Gray areas represent grid points where a single slope provides
  a better fit in terms of BIC.}
     \label{TippingPoints}
\end{figure}

\begin{figure*}
\centering
  \includegraphics[scale=0.7]{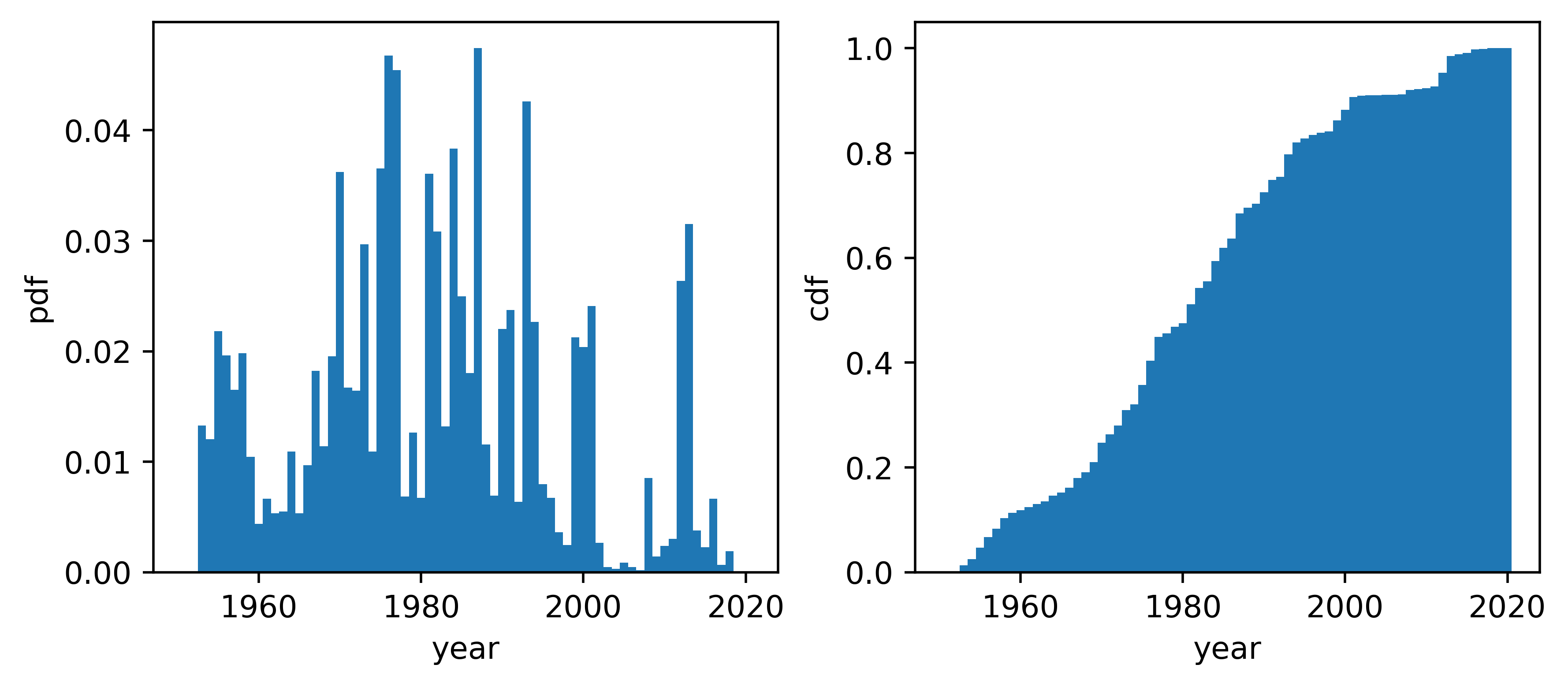}
  \caption{\label{Fig_changeyears}
    Left: A histogram of the number of grid points
    for which the change point lies in the respective year. The large
    fluctuations are stronger than expected for purely statistical reasons and
    hence seem to be related to natural climate variability on larger spatial
    scales. Right: the corresponding empirical cumulative distribution.}
\end{figure*}

The times of change in terms of the minimal RMS-error for all land grid points
are presented in Fig.\ref{TippingPoints}. We masked those grid points preferring
a single line fit with a gray color, since we do not consider the years of
change for these grid-points to carry meaningful information.
First of all, the figure shows that the change points 
in different regions have occurred at quite different times, from the early
years of the changing time interval (1960, 2010) to its latest.
This diversity can be seen in each continent. Nevertheless, more than 50\% of the changes occur in the years
between 1970s and 1990s, see
Fig.\ref{Fig_changeyears}. The very same figure also shows the huge
fluctuations from year to year, which are much larger than statistical
estimation errors (see Fig.\ref{Fig_synthetic})
and hence have a climatological meaning: Particularly low
annual temperatures tend to favor the change of the slope, see, e.g., 
Fig.\ref{global_data}, where the optimal change point is at the global minimum
of the temperature curve. A more detailed understanding of these fluctuations
is still lacking, and we found no good correlation with known oscillation
phenomena such as ENSO. 

From  Fig.\ref{TippingPoints} we also see interesting geographical patterns.
The continents Asia, North America, south of South America and south of Africa have gone
through a changing trend in the 1970s or they have remained in a constant
trend according to Fig. \ref{dBIC}.
The north of North America, Europe, the
Middle East, North Africa, northern South America, and Australia underwent
a change of their warming trend in the 1980s and 1990s. At latest,
Siberia, Alaska, and the west of Australia have observed changing trends after 2000.

Although diversity in the times of change is not unexpected, a closer look 
at Fig.\ref{TippingPoints} raises a question about why there are regions of
early tipping in the direct vicinity of regions with rather late tipping. For example, Fig.\ref{TippingPoints} 
exhibits that some region in western Australia 
has changed to the new trend in the 2000s but is close to a region that
has experienced a changing trend before the 1980s. Similar patterns are observed in
other regions of the globe as well. While the implications for understanding climate change
are less clear, our analysis can at least help us finding the statistical
cause of these abrupt changes of the change points as a function of spatial position: 
The local geographic attributes of climate change related temperature
increase are superimposed by natural climate variability with much larger
fluctuations than the global mean temperature.
These variations can be traced in
the behavior of the RMS-error curves as a function of the year of change of
the two slopes in our fits. Indeed, our survey reveals that the
uniqueness of a minimum of the RMS-error can not be guaranteed
because of the complicated temperature variations. In other words, variations
in the temperature time series for many grid points lead to the existence of
multiple relative minima, of which we chose the absolute one as the optimal
time of change (Fig. \ref{global_data} provides a typical
example). However, if we compare data on neighboring grid points, their time
series are similar due to spatial correlations of temperature variations,
and therefore also their error curves are similar with
minima occurring in the same (or adjacent) years.
Despite this, there can be a jump of the optimal change point, simply
because the depth of the minima changes and hence the absolute minima of
neighboring grid points are in
different years. We have to admit that this is some weakness of the method,
and we will pick up this discussion again in the Conclusions.

It should be mentioned that our investigation shows that Antarctica and
the Northern Hemisphere, those regions whose deterministic dynamics are
identified as a single-segment, are the regions with the strongest short-term
variations over this time interval.

\begin{figure}
	\centering
%	\begin{subfigure}{\linewidth}
%		\includegraphics[scale=0.75]{Map__Slop_of_line1_wide_masked.png}\\
		\includegraphics[scale=0.6]{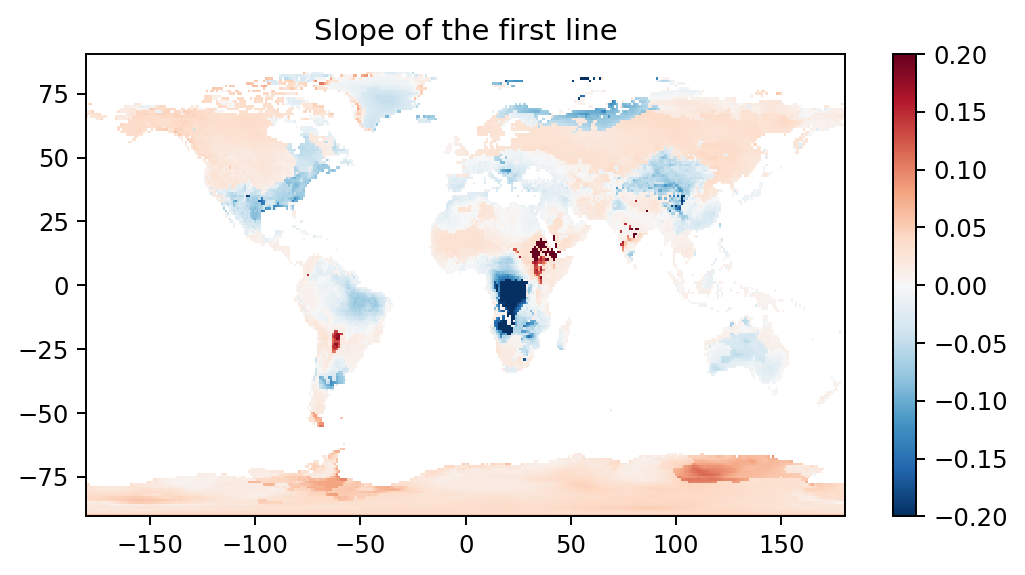}\\
		%\caption{}
%		\label{slope1}
%	\end{subfigure}
%	\begin{subfigure}{\linewidth}
%		\includegraphics[scale=0.75]{Map__Slop_of_line2_wide_masked.png}\\
		\includegraphics[scale=0.6]{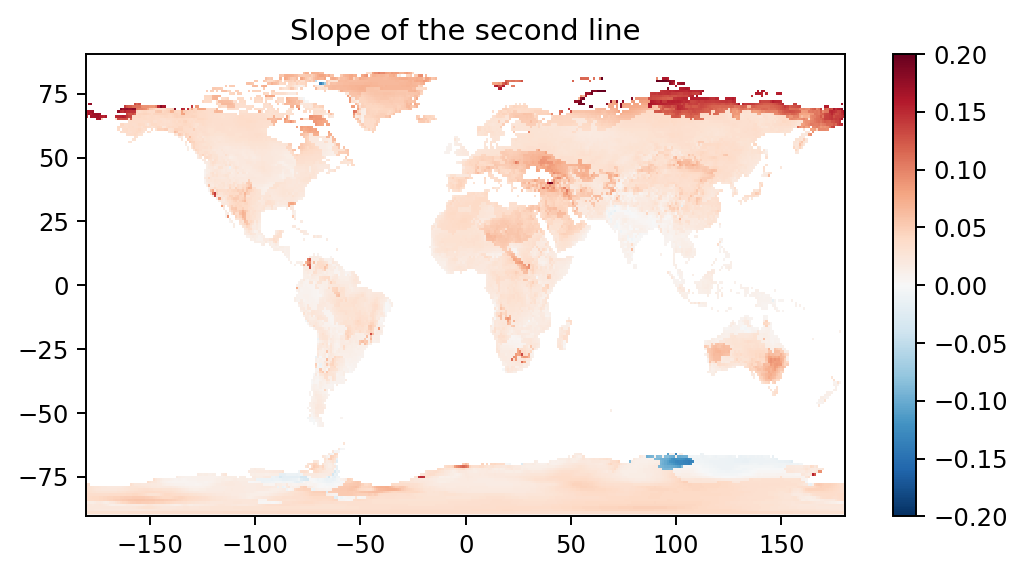}\\
		%\caption{}
%		\label{slope2}
%	\end{subfigure}
%	\begin{subfigure}{\linewidth}
%	        \includegraphics[scale=0.75]{Map__SlopeDifference_line2frome1_wide_masked.png}
	        \includegraphics[scale=0.6]{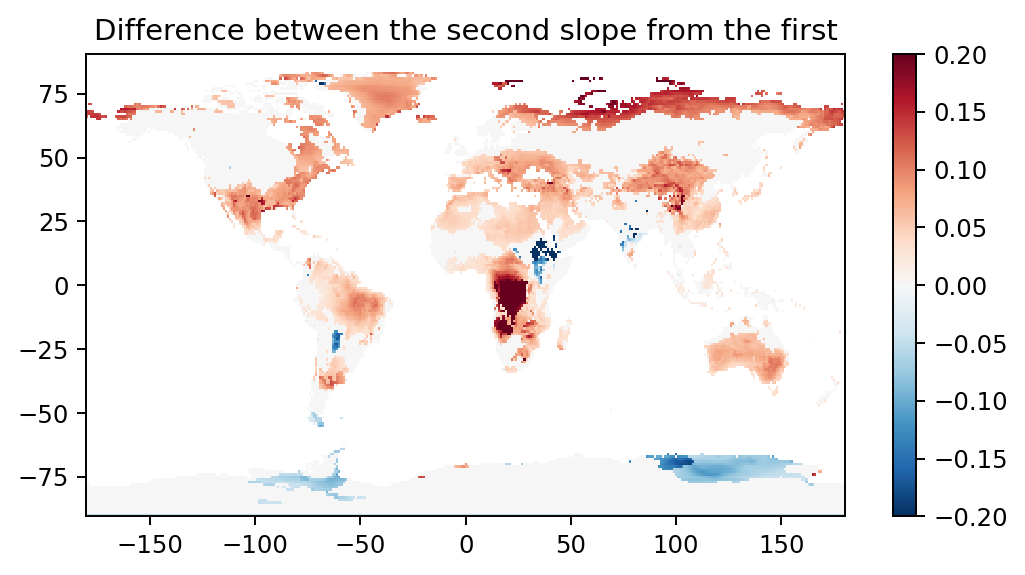}
	        %\caption{}
%	        \label{diff_slope2_slope1}
%         \end{subfigure}
\caption{(a) The slope of the first fitted line in $K/year$. In every
  continent we find both slopes with negative and positive values.
  The center of North America and the south of Africa exhibit 
  the highest and the lowest slopes, shown by bold colors. (b)
  The slope of the second part of the fit. The slopes are mostly positive confirming the
  warming trend. (c) The difference between the second and the first slope.
  For those grid points where BIC suggests a single linear fit, we show its
  slope in both (a) and (b), while the value is 0 in (c).  
  \label{fig:trend_maps}} 
	\label{slopes}
\end{figure}

In addition to \textit{when} the trend changed, we present in 
Fig.\ref{fig:trend_maps} \textit{how} the trend changed,
by showing in a color-code a map of the slopes of the first and of the second linear segment of the optimal fits, and in panel (c) also their differences.
Fig.\ref{fig:trend_maps}(b) confirms the well-known climate change
in almost the whole globe by showing mostly positive values in the range of
$0.03~Kelvin/year~(3.0~Kelvin/century)$. We should note that those 
regions in Figure \ref{fig:trend_maps}(c) where the differences are largest
mostly correspond to areas where the change 
point occurred early or late. These regions require additional study to
explore the trend differences with high precision. Consequently, we tend to
focus on other regions of the globe for our analysis. 

In general, as suggested by Figure \ref{fig:trend_maps}, the majority of local
trends before the change points are non-zero, displaying both cooling and
warming trends across various regions, resulting in a distribution of trend
values around zero. However, after the change point, there is a notable
transition towards positive trend values. Thus, local regions have experienced
different scenarios. 

For example, the red color on the maps Figs.\ref{fig:trend_maps}(a) and
(b) in Siberia, north of Africa, north of Australia is the signature of the
warming trend lasting for more than 70 years. However, after the change point
the trend increased, meaning an acceleration of warming. 
Interestingly, there are also regions in Antarctica and South America where the
trend values have been positive both before and after the change
point but where the values of the second slope are smaller,
indicating a transition from the earlier intensive warming phase to a
more moderate one. 

Another interesting set of regions includes southern North America, northern
South 
America, China, and South Africa, as indicated by the blue color on map
Fig. \ref{fig:trend_maps}(a) and red color in (b) and (c). In these areas,
the trends switch from a cooling
to a warming phase, signaling significant climate change. The change from
negative to positive trends in these areas, predominantly occurring around the
1970s as illustrated in Fig. \ref{TippingPoints}, underscores the predominance
of global warming over local trends. 

Europe, western Asia, and North Africa exhibit characteristics of well-known
"global change", defined as a transition from a relatively flat slope to a
positive one, as established in the literature. Although these regions
experience pronounced warming with a delay (see Fig. \ref{TippingPoints}),
this pattern suggests that they are prime candidates for investigating climate
change through the analysis of relevant time series data.

Besides the detailed information discussed above, Fig\ref{fig:trend_maps}
depicts two
significant facts about global warming. First, warming speed does not occur
homogeneously on the globe but in the southern hemisphere slower than in the
north. A possible reason might be the much larger water mass in the southern
hemisphere with its huge heat capacity.

The second interesting feature that Figs.\ref{fig:trend_maps}
reveal is
about the rare regions in which the temperature trend decreased which is
contrary to the expectation. Exploring the reason for this 
paradox requires further geographical, climatological, and environmental
studies beyond this work's scope. Still, we believe it can open
new gates toward strategies for controlling global warming.

\section{Conclusions}
In this paper, we argue that the variation of local temperatures over the past
70 years is significantly more complex than what can be accounted for by a
simple linear trend model. 
In the hierarchy of model complexity and with a desire for analytically
tractable models, the next parsimonious model is a model of two linear
segments that merge continuously. This model has 4 free parameters to be
adapted to the data: 3 parameters for the two slopes and one off-set, and the
time of the change from one slope to the other. We fix all of these with a
global least square fit, where we first fix the time for the change point, then
solve the least squares problem to determine the 3 parameters of this model,
and eventually, minimize the RMS error with respect to the time of change by
repeating this fit for all possible change points.
We show as results maps of the optimal times of change, of the two slopes, and
of their differences. We compared the statistical significance of such a model
opposed to a single trend fit by rejection of the null hypothesis at 95\%
confidence level and used the Bayesian information criterion BIC for model
selection.

\begin{figure*}
  \centerline{\includegraphics[scale=0.6]{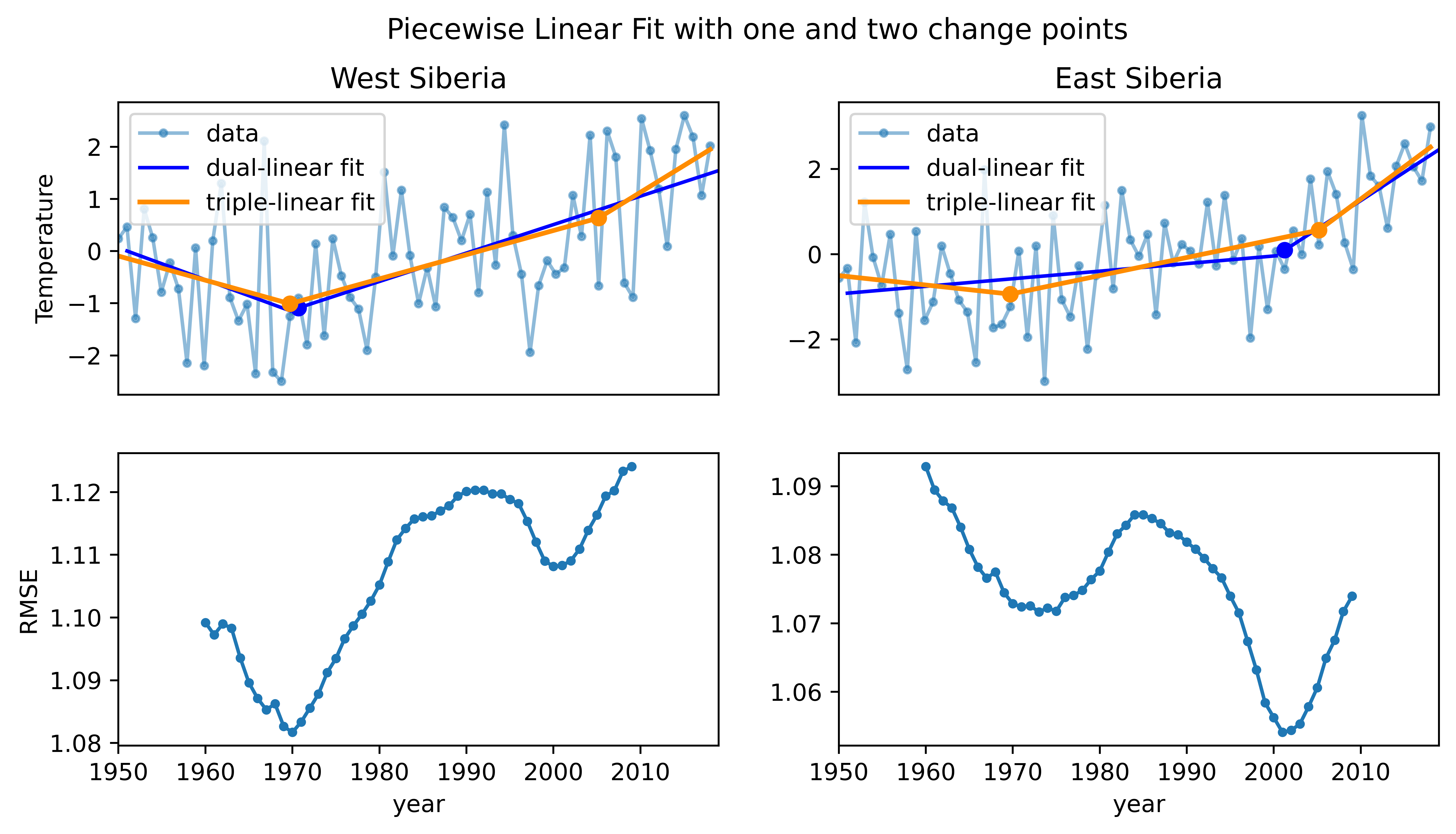}}
  \caption{\label{fig:Siberia}
Raw data and fits for one grid point each 
    in the west and the east of Siberia, where we detected two very different
    years of change. The RMSE curves of our dual-linear fits as function of
    where we place the time of change shows two minima at almost equal times 
    for both data sets, where the absolute minimum is used to determine the
    year of change. In this case, a fit with a three slope model is more
    robust and might be the more appropriate model.}
  \end{figure*}

Our main conclusion is that such a local analysis reveals many relevant and
interesting features of how climate change takes place locally. While we see
that temperature increase accelerates in many regions of the world, there are
some where it slows down and even a few areas where the temperature gets colder.
The regions with an increase of the temperature trend are in particular the
northern land masses of the northern hemisphere, which, e.g., is a bad message
for the permafrost ground in Siberia (methane release)
and for the Greenland ice sheet (sea level rise). 
For the Antarctic our results are inconclusive. This is, however, in
line with other recents studies: The warming of the Antartic land masses
has been discussed controversally \cite{Antarctica1,Antarctica2, Antarctica3}.
\\
Our analysis also show some at first sight strange
features. Fig.\ref{TippingPoints} 
reveals that we detect a change of the warming trend in west Siberia in the
1970s, where in eastern part of Siberia it occurs in late 1990s. The reason
for this switch is that Siberian data show two distinct relative minima of the
RMS error, the
early one being the absolute minumum in the west, the later one in the east.
Fig.\ref{fig:Siberia} shows this together with the raw data, our two-slopes
fits, and fits with three linear segments. Evindently, for this part of the
globe, a model with two change points might be adequate and more robust.
As said before, such a model is beyond the scope of this paper, but we will
devote forthcoming work to extensions of the model and the issue of
model-misspecification. Technically speaking, there are no change points in
temperature series, nobody has turned a switch at a given time and thereby
changed the warming trend. All these models are just approximations to the
complex time evolution and they must prove their usefulness by the way we can
draw conclusions from such an analysis.

Evidently, since our
results are a straightforward statistical analysis of data, these can only be
as good as the data. It is known that re-analysis data might suffer from the
lack of observation data in the pre-satellite era in less populated regions
of the globe. From this point of view, it would be safer to start this analysis
with the year 1979, when remote sensing data from satellites
entered the data
assimilation. However, since the global mean surface temperature shows a
change point in the 7th decade, we decided to work with time series starting in
1950, i.e., as far back as the ERA5 data set reaches back. 

Our numerical analysis falls into the set of statistical methods known as
change point detection. Nonetheless, as we discussed already in the
introduction, one can also interpret it in the context of tipping points in
the climate system. As it has been discussed by many authors, there are
several components in the Earth system, where temperature increase triggers
the onset of a feedback loop that eventually will accelerate temperature
increase, such as the sea ice-albedo effect in the Arctic, or methane release
from permafrost grounds. Therefore, our analysis can also be interpreted in
terms of tipping. In those regions of the globe, where the second slope is
considerably larger than the first slope (red areas on the map
Fig.\ref{fig:trend_maps} (c), one can speculate whether they have passed a
tipping point of their local climate already in the past.

\begin{appendices}

\section{}\label{secA1}

We use the Lagrangian optimization method to find the optimal slopes of the
two segments under the constraint that the lines merge in the given point
$T$. Since the optimal fit is determined by the minimum of the variance, the
Lagrangian function is as follows: 
\begin{equation}
\begin{split}
\mathcal{L}  &= \sum_{t=1}^{T}\left(x_t-(a_1 t+ b_1) \right)^2 + \sum_{t=T+1}^{N}\left(x_t-(a_2 t+ b_2) \right)^2 \\
& + \lambda \left( a_1 T+ b_1 -(a_2 T + b_2)\right)
\label{Eq_AppA1}
\end{split}
\end{equation}
where $\mathcal{L}$ is the Lagrangian function, $x_t$ is variable at the time $t$, $T$ is the merge point, $a_1$ and $a_2$ are the slopes of the fitted lines in the first and second segments, $b_1$ and $b_2$ are their intercepts, and $\lambda$ is the Lagrange multiplier. For numerical simplicity, we replace $t$ by $t^\prime=t-T$ in the second term:\\
\begin{equation}\label{Eq_AppA2}
\begin{split}
\mathcal{L}  &= \sum_{t=1}^{T}\left(x_t-(a_1 t+ b_1) \right)^2 + \sum_{t=1}^{N-T}\left(x_{t+T}-(a_2 t+ b_2) \right)^2\\
&
+ \lambda \left( a_1 T+ b_1 - b_2\right),
\end{split}
\end{equation}
The slopes are determined by finding the stationary states of 
$\mathcal{L}$ as a function of Lagrange parameters:
\begin{equation}\label{Eq_AppA3}
\begin{split}
\frac{\partial \mathcal{L}}{\partial a_1} &= -2\sum_{t=1}^{T}\left(x_t-(a_1 t+ b_1) \right)t + \lambda T =0\\
\frac{\partial \mathcal{L}}{\partial b_1} &= -2\sum_{t=1}^{T}\left(x_t-(a_2 t+ b_2) \right)+ \lambda=0\\
\frac{\partial \mathcal{L}}{\partial a_2} &= -2\sum_{t=1}^{N-T}\left(x_{t+T}-(a_2 t+ b_2)\right)t=0\\
\frac{\partial \mathcal{L}}{\partial b_2} &= -2\sum_{t=1}^{N-T}\left(x_{t+T}-(a_2 t+ b_2) \right)- \lambda=0\\
\frac{\partial \mathcal{L}}{\partial \lambda} &= a_1 T +b_1 - b_2 =0 
\end{split}
\end{equation}
which can be rewritten as:
\begin{equation}\label{Eq_AppA4}
\begin{split}
& (XT)_1 - a_1\sum_{t=1}^T t^2 - b_1\sum_{t=1}^T t + \tilde{\lambda} T = 0\\
& X_1 - a_1\sum_{t=1}^T t - b_1 T + \tilde{\lambda} = 0\\
& (XT)_2 - a_2\sum_{t=1}^{N-T} t^2 - b_2\sum_{t=1}^{N-T} t = 0\\
& X_2 - a_2\sum_{t=1}^{N-T} t - b_2(N-T)- \tilde{\lambda} = 0\\
& a_1 T +b_1 - b_2 =0 
\end{split}
\end{equation}
where $\tilde{\lambda} = -(1/2)\lambda$, $(XT)_1= \sum_{t=1}^{T} t x_t$, $X_1= \sum_{t=1}^{T} x_t$, $(XT)_2= \sum_{t=1}^{N-T} t x_{t+T}$ and $X_2= \sum_{t=1}^{N-T} x_{t+T}$. \\
Since $\sum_{t=1}^T t = T(T+1)/2$, $\sum_{t=1}^T t^2 = T (T+1)(2T+1)/6$,  $\sum_{t=1}^{N-T} t = (N-T)(N-T+1)/2$ and $\sum_{t=1}^{N-T} t^2 = (N-T)(N-T+1)(2(N-T)+1)/6$ are known based on the $N$ and $T$, the set of equations are simplified:\\
\begin{equation}\label{Eq_AppA5}
\begin{split}
& (XT)_1 - a_1\frac{T(T+1)(T+1/2)}{3} - b_1\frac{T(T+1)}{2} + \tilde{\lambda}T = 0\\
& X_1 - a_1\frac{T(T+1)}{2}- b_1 T + \tilde{\lambda} = 0\\
& (XT)_2 - a_2\frac{(N-T)(N-T+1)(N-T+1/2)}{3}\\
& - b_2\frac{(N-T)(N-T+1)}{2}= 0\\
& X_2 - a_2\frac{(N-T)(N-T+1)}{2} - b_2 (N-T) - \tilde{\lambda} = 0\\
& a_1 T +b_1 - b_2 =0 
\end{split}
\end{equation}

Solving this set of equations, we obtain the formulae for the coefficients $a_1$, $b_1$, $a_2$, and $b_2$:
\begin{equation}\label{Eq_AppA6}
\begin{split}
a_1 =& N^{-1}(2 N T^3 - 3 N T^2 + N T - 2 T^4 + 4 T^3 - T^2 - T)^{-1}\\
&(-6 N^2 T X_1 + 6 N^2(XT)_1 + 12 N T^2 X_2 - 6 N T X_1 \\
&- 12 N T X_2 + 12 N T (XT)_1 - 6 N (XT)_1 + 6 T^3 X_1 \\
&- 12 T^3 X_2 + 18 T^2 X_2 - 18 T^2 (XT)_1 - 18 T^2 (XT)_2\\
&- 6 T X_1 - 6 T X_2 + 18 T (XT)_1 + 18 T (XT)_2)
\end{split}
\end{equation}
\begin{eqnarray}\label{Eq_AppA7}
\begin{split}
a_2 =& N^{-1}(2 N^3 T - N^3 - 6 N^2 T^2 + 6 N^2 T + 6 N T^3\\
&- 9 N T^2 + N T + N - 2 T^4 + 4 T^3 -T^2 -T)^{-1}\\
& (6N^2 T X_1 - 12 N^2 T X_2 + 6 N^2 X_1 + 6 N^2 X_2 \\
&- 18 N^2 (XT)_1 - 12 N T^2 X_1 + 24 N T^2 X_2 - 6 N T X_1\\
&- 24 N T X_2 + 36 N T (XT)_1 + 24 N T (XT)_2 + 6 N X_1\\
&+ 6 N X_2 - 18 N (XT)_1 - 12 N (XT)_2 + 6 T^3 X_1\\
&- 12 T^3*X_2 + 18 T^2 X_2 - 18 T^2 (XT)_1- 18 T^2 (XT)_2 \\
&- 6 T X_1 - 6 T X_2 + 18 T (XT)_1 + 18 T (XT)_2)
\end{split}
\end{eqnarray}
\begin{equation}\label{Eq_AppA8}
\begin{split}
b_1 =& N^{-1}(2 N T^2 - 3 N T + N - 2 T^3 + 4 T 2 - T - 1)^{-1}\\
&(6 N^2 T X_1 - 6 N^2 (XT)_1 - 4 N T^2 X_1 - 4 N T^2 X_2 \\
&+ 6 N T X_1 + 4 N X_1 + 4 N X_2 - 6 N (XT)_1 - 2 T^3 X_1 \\
&+ 4 T^3 X_2 - 2 T^2 X_1 - 2 T^2 X_2 + 6 T^2 (XT)_1 + 6 T^2 (XT)_2\\
&+ 2 T X_1 - 4 T X_2 + 2 X_1 + 2 X_2 - 6 (XT)_1 - 6 (XT)_2)
\end{split}
\end{equation}
\begin{equation}\label{Eq_AppA9}
\begin{split}
b_2 =& N^{-1}(2 N T - N - 2 T^2 + 2 T + 1)^{-1}\\
&(-4.0 N T X_1 + 8 N T X_2 - 4 N X_1 - 4 N X_2 + 12 N (XT)_1 \\
&+ 4 T^2 X_1 - 8 T^2 X_2 + 2 T X_1 + 8 T X_2 - 12 T (XT)_1 \\
&- 12 T (XT)_2 - 2 X_1 - 2 X_2 + 6 (XT)_1 + 6 (XT)_2)
\end{split}
\end{equation}

Therefore, the coefficients are obtained by calculating  $(XT)_1$ and $(XT)_2$
for a given time series and changing point. 

\end{appendices}

\bibliographystyle{plain}       % Choose the style you want
\newpage
\newpage
\bibliography{bibtex}

\begin{thebibliography}{10}

\bibitem{AIC}
H.~Akaike.
\newblock A new look at the statistical model identification.
\newblock {\em IEEE Transactions on Automatic Control}, 19:716--723, 1974.

\bibitem{review_change_points}
S.~Aminikhanghahi and D.~J. Cook.
\newblock A survey of methods for time series change point detection.
\newblock {\em Knowledge and Information Systems}, 51:339--367, 2017.

\bibitem{ModelSelection2}
D.~Anderson and K.~Burnham.
\newblock {\em Model Selection and Multi-model Inference}.
\newblock Springer-Verlag, second edition, 2004.

\bibitem{Arrhenius}
S.~Arrhenius.
\newblock Xxxi. on the influence of carbonic acid in the air upon the
  temperature of the ground.
\newblock {\em Philosophical Magazine and Journal of Science}, 41:237--276,
  1896.

\bibitem{Antarctica2}
D.~Bromwich, J.~Nicolas, A.~Monaghan, M.~Lazzara, L.~Keller, G.~Weidner, and
  A.~Wilson.
\newblock Erratum: Corrigendum: Central west antarctica among the most rapidly
  warming regions on earth.
\newblock {\em Nature Geoscience}, 7:76--76, 2014.

\bibitem{weighting_grid_points}
K.~Cowtan, P.~Jacobs, P.~Thorne, and R.~Wilkinson.
\newblock Statistical analysis of coverage error in simple global temperature
  estimators.
\newblock {\em Dynamics and Statistics of the Climate System}, 3, 2018.

\bibitem{CDC1}
{Deutscher Wetterdienst}.
\newblock Climate data center of the german weather service.
\newblock \url{https://www.dwd.de}.
\newblock Usage under CC BY 4.0. Potsdam annual mean temperatures downloaded
  from
  \url{https://opendata.dwd.de/climate_environment/CDC/observations\_germany/climate/annual/kl/historical/},
  station No. 03987, data set
  jahreswerte\_KL\_03987\_18930101\_20241231\_hist.zip containing data from
  1893 till 2024.

\bibitem{report_2024_DWD}
{Deutscher Wetterdienst}.
\newblock Klimastatusbericht 2024.
\newblock
  \url{https://www.dwd.de/DE/leistungen/klimastatusbericht/publikationen/ksb_2024.pdf},
  2024.
\newblock (available only in German).

\bibitem{DingTarokhYang}
J.~Ding, V.~Tarokh, and Y.~Yang.
\newblock Model selection techniques: An overview.
\newblock {\em IEEE Signal Processing Magazine}, 35:16--34, 2018.

\bibitem{Bundeetal}
J.~F. Eichner, E.~Koscielny-Bunde, A.~Bunde, S.~Havlin, and H.-J. Schellnhuber.
\newblock Power-law persistence and trends in the atmosphere: a detailed study
  of long temperature records.
\newblock {\em Physical Review E}, 68:046133, 2003.

\bibitem{Inhomogeneity1}
F.~Estrada, D.~Kim, and P.~Perron.
\newblock Spatial variations in the warming trend and the transition to more
  severe weather in midlatitudes.
\newblock {\em Scientific Reports}, 11:145, 2021.

\bibitem{ECMWF}
{European Centre for Medium-Range Weather Forecasts (ECMWF)}.
\newblock The family of era5 datasets.
\newblock
  \url{https://confluence.ecmwf.int/display/CKB/The+family+of+ERA5+datasets}.
\newblock as viewed in January 2025.

\bibitem{specialregion1}
M.~Falvey and R.~Garreaud.
\newblock Regional cooling in a warming world: Recent temperature trends in the
  southeast pacific and along the west coast of subtropical south america
  (1979–2006).
\newblock {\em Journal of Geophysical Research}, 114, 2009.

\bibitem{ERA5_2}
H.~Hersbach, B.~Bell, P.~Berrisford, G.~Biavati, A.~Horányi,
  J.~Muñoz-Sabater, J.~Nicolas, C.~Peubey, R.~Radu, I.~Rozum, D.~Schepers,
  A.~Simmons, C.~Soci, D.~Dee, and J.~Thépaut.
\newblock Era5 hourly data on single levels from 1940 to present, 2023.
\newblock Copernicus Climate Change Service (C3S) Climate Data Store (CDS), as
  viewed in January 2025.

\bibitem{era5_reanalysisProcess}
H.~Hersbach, B.~Bell, P.~Berrisford, S.~Hirahara, A.~Horányi,
  J.~Muñoz-Sabater, J.~Nicolas, C.~Peubey, R.~Radu, D.~Schepers, et~al.
\newblock The era5 global reanalysis.
\newblock {\em Quarterly Journal of the Royal Meteorological Society},
  146:1999--2049, 2020.

\bibitem{IPCC}
{IPCC}.
\newblock Intergovernmental panel on climate change.
\newblock \url{https://www.ipcc.ch}.
\newblock Contains links to all reports.

\bibitem{Johannes}
J.~Kassel and H.~Kantz.
\newblock Statistical inference of one-dimensional persistent nonlinear time
  series and application to predictions.
\newblock {\em Physical Review Research}, 4:013206, 2022.

\bibitem{Koscielny}
E.~Koscielny-Bunde, A.~Bunde, and S.~Havlin.
\newblock Analysis of daily temperature fluctuations.
\newblock {\em Physica A: Statistical Mechanics and its Applications},
  231:393--396, 1996.

\bibitem{KoscielnyBunde}
E.~Koscielny-Bunde, H.~E. Roman, A.~Bunde, and S.~Havlin.
\newblock Long-range power-law correlations in local daily temperature
  fluctuations.
\newblock {\em Philosophical Magazine B}, 77:1331--1340, 1998.

\bibitem{tipping_precursor}
T.~Lenton.
\newblock Early warning of climate tipping points.
\newblock {\em Nature Climate Change}, 1:201--209, 2011.

\bibitem{Lentonetal}
T.~Lenton, H.~Held, E.~Kriegler, J.~Hall, W.~Lucht, S.~Rahmstorf, and
  H.~Schellnhuber.
\newblock Tipping elements in the earth's climate system.
\newblock {\em Proceedings of the National Academy of Sciences},
  105:1786--1793, 2008.

\bibitem{Inhomogeneity3}
I.~Mahlstein, R.~Knutti, S.~Solomon, and R.~Portmann.
\newblock Early onset of significant local warming in low latitude countries.
\newblock {\em Environmental Research Letters}, 6:034009, 2011.

\bibitem{Potsdam_data}
M.~Massah and H.~Kantz.
\newblock Confidence intervals for time averages in the presence of
  long‐range correlations, a case study on earth surface temperature
  anomalies.
\newblock {\em Geophysical Research Letters}, 43:9243--9249, 2016.

\bibitem{review1}
M.~Mudelsee.
\newblock Trend analysis of climate time series: A review of methods.
\newblock {\em Earth-Science Reviews}, 190:310--322, 2019.

\bibitem{globalmeantemp}
{Multiple Sources}.
\newblock Global mean surface temperature records.
\newblock \url{https://en.wikipedia.org/wiki/Global_surface_temperature},
  \url{https://gml.noaa.gov/data/data.php},
  \url{https://crudata.uea.ac.uk/cru/data/temperature/},
  \url{https://data.giss.nasa.gov/gistemp/},
  \url{https://www.nsstc.uah.edu/data/msu/t2lt/}.
\newblock as viewed in January 2025.

\bibitem{ERA5}
Joaquín Muñoz-Sabater, Emanuel Dutra, Anna Agustí-Panareda, Clément
  Albergel, Giacomo Arduini, and Gianpaolo Balsamo.
\newblock Era5-land: A state-of-the-art global reanalysis dataset for land
  applications.
\newblock {\em Earth System Science Data}, 13:4349--4383, 2021.

\bibitem{nasa}
{NASA}.
\newblock Global temperature vital signs.
\newblock \url{https://climate.nasa.gov/vital-signs/global-temperature/}, 2024.

\bibitem{impact1}
J.~Patz, D.~Campbell-Lendrum, T.~Holloway, and J.~Foley.
\newblock Impact of regional climate change on human health.
\newblock {\em Nature}, 438:310--317, 2005.

\bibitem{DFA}
C.~Peng, S.~Buldyrev, S.~Havlin, M.~Simons, H.~Stanley, and A.~Goldberger.
\newblock Mosaic organization of dna nucleotides.
\newblock {\em Physical Review E}, 49:1685--1689, 1994.

\bibitem{Iwanetal}
E.~Phillips, M.~Höll, H.~Kantz, and Y.~Zhou.
\newblock Trend analysis in the presence of short-and long-range correlations
  with application to regional warming.
\newblock {\em Physical Review E}, 108, 2023.

\bibitem{Rahmstorf}
S.~Rahmstorf, G.~Foster, and N.~Cahill.
\newblock Global temperature evolution: recent trends and some pitfalls.
\newblock {\em Environmental Research Letters}, 12:054001, 2017.

\bibitem{Antarctica1}
G.~Retamales-Muñoz, C.~Durán-Alarcón, and C.~Mattar.
\newblock Recent land surface temperature patterns in antarctica using
  satellite and reanalysis data.
\newblock {\em Journal of South American Earth Sciences}, 95:102304, 2019.

\bibitem{BIC}
G.~Schwarz.
\newblock Estimating the dimension of a model.
\newblock {\em Annals of Statistics}, 6:461--464, 1978.

\bibitem{Antarctica3}
E.~Steig, D.~Schneider, S.~Rutherford, M.~Mann, J.~Comiso, and D.~Shindell.
\newblock Warming of the antarctic ice-sheet surface since the 1957
  international geophysical year.
\newblock {\em Nature}, 457:459--462, 2009.

\bibitem{ModelSelection}
P.~Stoica and Y.~Selen.
\newblock Model-order selection: a review of information criterion rules.
\newblock {\em IEEE Signal Processing Magazine}, 21:36--47, 2004.

\bibitem{Stone}
M.~Stone.
\newblock An asymptotic equivalence of choice of model by cross-validation and
  akaike's criterion.
\newblock {\em Journal of the Royal Statistical Society. Series B
  (Methodological)}, 39:44--47, 1977.

\bibitem{Inhomogeneity2}
D.~You-Ren, H.~Bjørn, and F.~Samset.
\newblock Evaluating global and regional land warming trends in the past
  decades with both modis and era5-land land surface temperature data.
\newblock {\em Remote Sensing of Environment}, 280, 2022.

\bibitem{NOAA_global}
H.~Zhang, B.~Huang, J.~Lawrimore, M.~Menne, and T.~Smith.
\newblock Noaa global surface temperature dataset (noaaglobaltemp), version
  4.0.
\newblock
  \url{https://www.ncei.noaa.gov/metadata/geoportal/rest/metadata/item/gov.noaa.ncdc%253AC01704/html},
  2019.
\newblock
  \url{https://www.ncei.noaa.gov/access/monitoring/climate-at-a-glance/global/time-series}.

\bibitem{impact2}
T.~Zhang and Y.~Huang.
\newblock Estimating the impacts of warming trends on wheat and maize in china
  from 1980 to 2008 based on county level data.
\newblock {\em International Journal of Climatology}, 33:699--708, 2013.

\end{thebibliography}

\end{document}